\documentstyle[epsf]{l-aa}           
\begin{document}

\thesaurus{06(08.22.2, 08.15.1, 08.05.3, 08.16.5)}
\title{Period changes of $\delta$ Scuti stars and stellar evolution}
\author{M.~Breger\inst{1} \and A. A. Pamyatnykh \inst{1,2,3}}

\offprints{M.~Breger}

\institute{Astronomisches Institut der Universit\"at Wien,
T\"urkenschanzstr. 17,
A--1180 Wien, Austria\\INTERNET: breger@astro.univie.ac.at
\and
Copernicus Astronomical Center, Bartycka 18, PL-00-716 Warsaw, Poland
\and
Institute of Astronomy, Russian Academy of Sciences,
Pyatnitskaya Str. 48, 109017 Moscow, Russia}

\date{Received date; accepted date}

\maketitle
\markboth{M. Breger \& A. A. Pamyatnykh : Period changes and evolution}
{M. Breger \& A. A. Pamyatnykh : Period changes and evolution}

\begin{abstract}

Period changes of $\delta$ Scuti stars have been collected
or redetermined from the available observations and are compared with
values computed from evolutionary models with and without
convective core overshooting.

For the radial pulsators of Pop. I,
the observations indicate (1/P) dP/dt values around
10$^{-7}$ year$^{-1}$
with equal distribution between period increases and decreases.
The evolutionary
models, on the other hand, predict that the vast majority should show
increasing periods. This increase should be a factor of about ten times
smaller than observed. For nonradial
$\delta$ Scuti pulsators of Pop. I, the discrepancies are even larger.
The behavior suggests that for these relatively unevolved stars
the rate of evolution cannot be deduced from the period changes.

The period changes of most Pop. II $\delta$ Scuti (SX Phe) stars are
characterized by sudden
jumps of the order of $\Delta P/P \sim$ 10$^{-6}$. However, at least
one star, BL Cam, shows
a large, continuous period increase. The variety of observed behavior
also seems to
exclude an evolutionary origin of the changes.

Model calculations show that the evolutionary period changes of pre-MS
$\delta$ Scuti stars are a factor of 10 to 100 larger than those of MS
stars.
Detailed studies of selected pre-MS $\delta$ Scuti stars are suggested.

\end{abstract}

\section{Introduction}

$\delta$ Scuti variables are short-period pulsators situated inside
the classical
instability strip on and near the main sequence (hereafter called MS).
They represent stars in the
hydrogen-burning stage with masses from 1.5 $M_{\odot}$ to
about 2.5 $M_{\odot}$
for the more evolved variables.

The majority of $\delta$ Scuti stars pulsates with a multitude of
nonradial p-- and
g--modes (often of mixed character) and low amplitudes near the
detection limit of 0.5 millimag.
A numerically small subgroup has
exceptionally low rotational velocities of $v \leq $ 30 km s$^{-1}$
with pulsation
properties resembling the classical cepheids and RR Lyrae stars:
radial pulsation
in the fundamental or first overtone mode with high amplitudes
between 0.3 and 1 mag.
In the past, the stars in this high-amplitude (HADS) subgroup were
sometimes referred to as Dwarf Cepheids
or AI Vel stars. They are observationally favored
for the study of period stability because of large amplitudes
and the presence of only a few excited modes (usually one or two).

The period changes caused by stellar evolution in and across the
Lower Instability Strip permit an observational test of stellar
evolution theory, provided that other physical reasons
for period changes can be excluded. The period -- mean
density relation for pulsating stars predicts a
period-luminosity-color-mass relation of the form

\[\log P = -0.3 M_{\rm bol} - 3 \log T_{\rm eff} - 0.5 \log M + \log Q
+ 12.708\]
where P is the radial period and $M$ is the stellar mass in solar units.
For the $\delta$ Scuti variables with known radial
pulsation, the observed coefficients are in almost exact agreement with
those predicted by the period--mean density relation (Breger 1980).

This confirmation of the theoretical relation is very important: since
the period-luminosity-relation holds for stars in different
evolutionary stages, it should also hold for a single star at different
evolutionary stages. For individual stars,
evolutionary period changes must occur, at least over long time scales.

An evolutionary change in $T_{\rm eff}$ and $M_{\rm bol}$ leads to a
period change of size

\[\frac{1}{P} \frac{dP}{dt} = -0.69\frac{dM_{\rm bol}}{dt}
 - \frac{3}{T_{\rm eff}} \frac{dT_{\rm eff}}{dt}
 + \frac{1}{Q} \frac{dQ}{dt},\]
where the coefficient of 0.69 has been derived by multiplying the
coefficient of 0.3
by $\ln$ 10 from the differentiation.

In the Lower Instability Strip, where the $\delta$ Scuti stars are
found, stellar evolution leads to increasing periods in the
vast majority of stars, with predicted increases of (1/P)dP/dt from
10$^{-10}$ year$^{-1}$ on the MS to 10$^{-7}$ year$^{-1}$ for the
longer-period evolved variables.
(More exact values will be computed in a later section of this paper.)
Such period changes should be observable.

There have been several comparisons between theoretical and observed
period changes. The careful compilation by Percy et al. (1980) led
to inconclusive results. They state that `the study of period changes
has had a long and sometimes dubious history' and find that `in most
cases is a source of confusion and frustration.'

Breger (1990a) had more data at his disposal and noted that for the
four evolved
$\delta$ Scuti stars of Pop. I the observed period decreases were
in contradiction to the
radius increases predicted from stellar evolution. Guzik \& Cox (1991)
have examined possible explanations for the period decreases seen in
evolved $\delta$ Scuti stars. While their models indeed predicted
smaller period increases, period decreases were not found. The fact
that all four evolved $\delta$ Scuti stars show period
decreases may conceivably be a reflection of an additional period
changing mechanism (leading to random changes) coupled with
small-number statistics. The recent results for the star AI Vel
(Walraven et al. 1992)
support such a view. They have
found that only one of the two radial frequencies shows a
period increase, while the second frequency was essentially constant.
The example of AI Vel shows that (slow) stellar evolution is not
the only mechanism to generate changes in the periods of pulsation.
This means that for an individual star the conversion of observed
period changes into stellar evolution rates (e.g. radius changes) has
to be applied with caution. The nonevolutionary period changes should
cancel out only for a larger group of stars in the same stage of
evolution.

An important update and analysis was provided by Rodriguez et al.
(1995), who confirmed the observed period decreases for evolved
$\delta$ Scuti stars. They also found
relatively large period increases in excess of the expected values
for the less evolved Pop. I $\delta$ Scuti variables and decreasing
periods for Pop. II variables. Furthermore, their study provided
further evidence for the earlier suspicions that the observed period
changes are not caused by evolution alone.

The present paper examines period changes
in $\delta$ Scuti stars from a theoretical point of view, provides
a review of the observational status, and compares the two.

\section{Predicted period changes due to stellar evolution}

\subsection{Standard evolutionary models and their oscillations}

The models of 1.5--2.5 $M_{\odot}$ in MS and post-MS evolutionary
stages were constructed using a standard stellar evolution code which
was developed in its main parts
by B.~Paczy\'nski, M.~Koz{\l}owski and R.~Sienkiewicz
(private communication). The same code was used in our recent studies
of FG~Virginis (Breger et al. 1995) and XX~Pyxidis
(Handler et al. 1997a,b; Pamyatnykh et al. 1997).
The computations were performed starting from
chemically uniform models on the Zero-Age Main Sequence (ZAMS),
assuming an initial hydrogen abundance $X=0.70$ and
a heavy element abundance $Z=0.02$.
The standard mixing-length theory of convection
with the mixing-length parameter $\alpha=1.0$ was used.
The choice of the mixing-length 
parameter has only a small effect on our models because they are too
hot to have an effective energy transfer by convection in the stellar
envelope. For the opacities, we used latest version of the OPAL tables
(Iglesias \& Rogers 1996) supplemented with the low--temperature data
of Alexander \& Ferguson (1994). In all computations the OPAL equation
of state was used (Rogers et al. 1996).

Some $\delta$ Scuti stars may be very young evolving towards the MS.
We did not compute pre-MS evolution explicitly. Instead
we used the evolutionary tracks computed by D'Antona \& Mazzitelli
(1994) for the `MLT Alexander' set
of input parameters\footnote
{
Mixing--Length Theory of convection with $\alpha = {\it l}/H_{\rm p}$
 = 1.4; OPAL opacities due to Iglesias \& Rogers (1993) supplemented
with low--temperature opacities due to Alexander \& Ferguson (1994);
initial hydrogen abundance $X=0.70$, heavy element abundance $Z=0.019$.
}, which are close to those used in our computations.

\begin{figure}
\epsfclipon
\epsfxsize=82mm
\epsffile{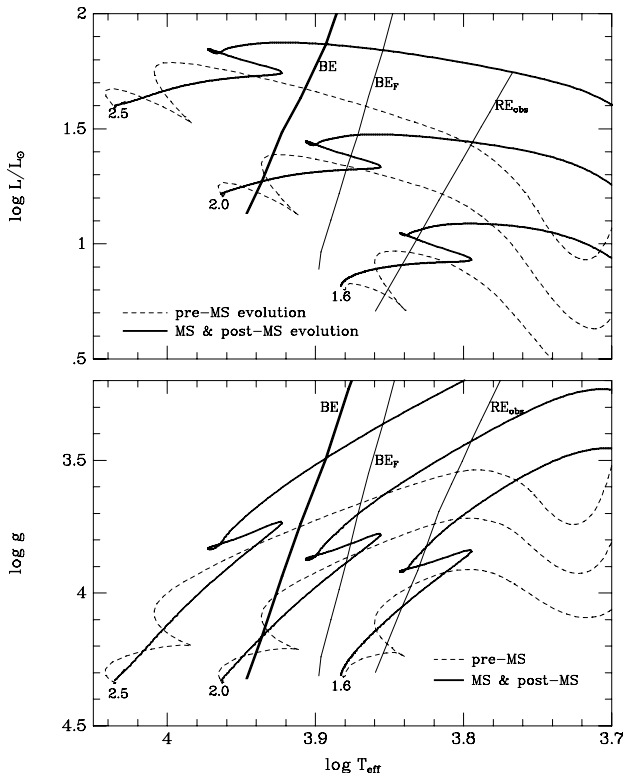}
\caption{Evolutionary tracks of 1.6, 2.0 and 2.5 $M_{\odot}$ models.
Solid lines correspond to our computations of MS and post-MS
evolution for initial hydrogen abundance $X=0.70$ and
heavy element abundance $Z=0.02$.
Dashed lines correspond to the pre-MS evolution as computed
by D'Antona \&
Mazzitelli (1994) for the `MLT Alexander' set of input parameters
(see text).
The symbols ${\rm BE}$, ${\rm BE_F}$ and ${\rm RE_{obs}}$
mark boundaries of the instability strip: the general theoretical
Blue Edge
for radial overtones, the theoretical Blue Edge for the radial
fundamental mode, and the empirical Red Edge. }
\end{figure}

The pre-MS and post-MS tracks are plotted in \mbox{Fig.~1}
for three values of stellar mass.
There is good agreement between the positions of ZAMS models
constructed by D'Antona \& Mazzitelli (1994) and our ZAMS models.
Therefore we can conclude that both
series of the computations are consistent with each other.

For MS and post-MS models we performed a linear nonadiabatic analysis
of low--degree oscillations ($\ell$\,$\leq$\,2) using the code
developed by W.~Dziembowski (for a general description see Dziembowski
1977). In Fig. 1 we have also shown the theoretical blue edges of the
$\delta$~Scuti instability strip which were computed for radial
oscillations of the
models with initial hydrogen abundance $X=0.70$ and heavy element
abundance $Z=0.02$. The general Blue Edge (${\rm BE}$) is the hottest
envelope of blue edges for overtones, from $p_8$ at the ZAMS
to $p_4$ in the upper parts of both panels.
There are no unstable modes in models to the left of this line.
${\rm BE_F}$ is the blue edge
for the radial fundamental mode. Note that in the region under
consideration the onset of the instability is practically insensitive
to the spherical degree of a mode, so that the position of the blue
edges will be unchanged when studying nonradial oscillations.
${\rm RE_{obs}}$ is the empirical Red Edge
transformed from the photometric data to the theoretical parameters
(we used simplest approximation about constancy of the convective flux
during the oscillation cycle, therefore we were not able to determine
theoretical red edge of the instability strip).

To study oscillations in the pre-MS evolutionary stage,
we constructed envelope models using global stellar
parameters (mass, effective temperature and luminosity)
along the evolutionary tracks computed by
D'Antona \& Mazzitelli (1994), and then computed radial oscillations
of these models. The envelopes were integrated from the surface
to a layer with temperature $\log T=7.2$ which corresponds to relative
radius $r/R=0.03-0.13$ in all models considered. For such deep
envelopes, the periods of radial oscillations are practically the same
as for full stellar models: in a 1.8 $M_{\odot}$ model near the ZAMS
where the errors are largest, the difference in the low--overtone
periods between envelope and full models does not exceed 0.07~\%.
These errors have no influence on the determination of evolutionary
period changes. To calculate period changes we used
one more result of D'Antona \& Mazzitelli: the age of the correspoding
pre-MS model.

Note that our evolutionary tracks and periods of oscillations
agree  with those from a recent
paper by Petersen \& Christensen--Dalsgaard (1996).

\subsection{Evolutionary period changes}

\begin{figure*}
\epsfclipon
\epsfxsize=180mm
\epsffile[41 158 496 632]{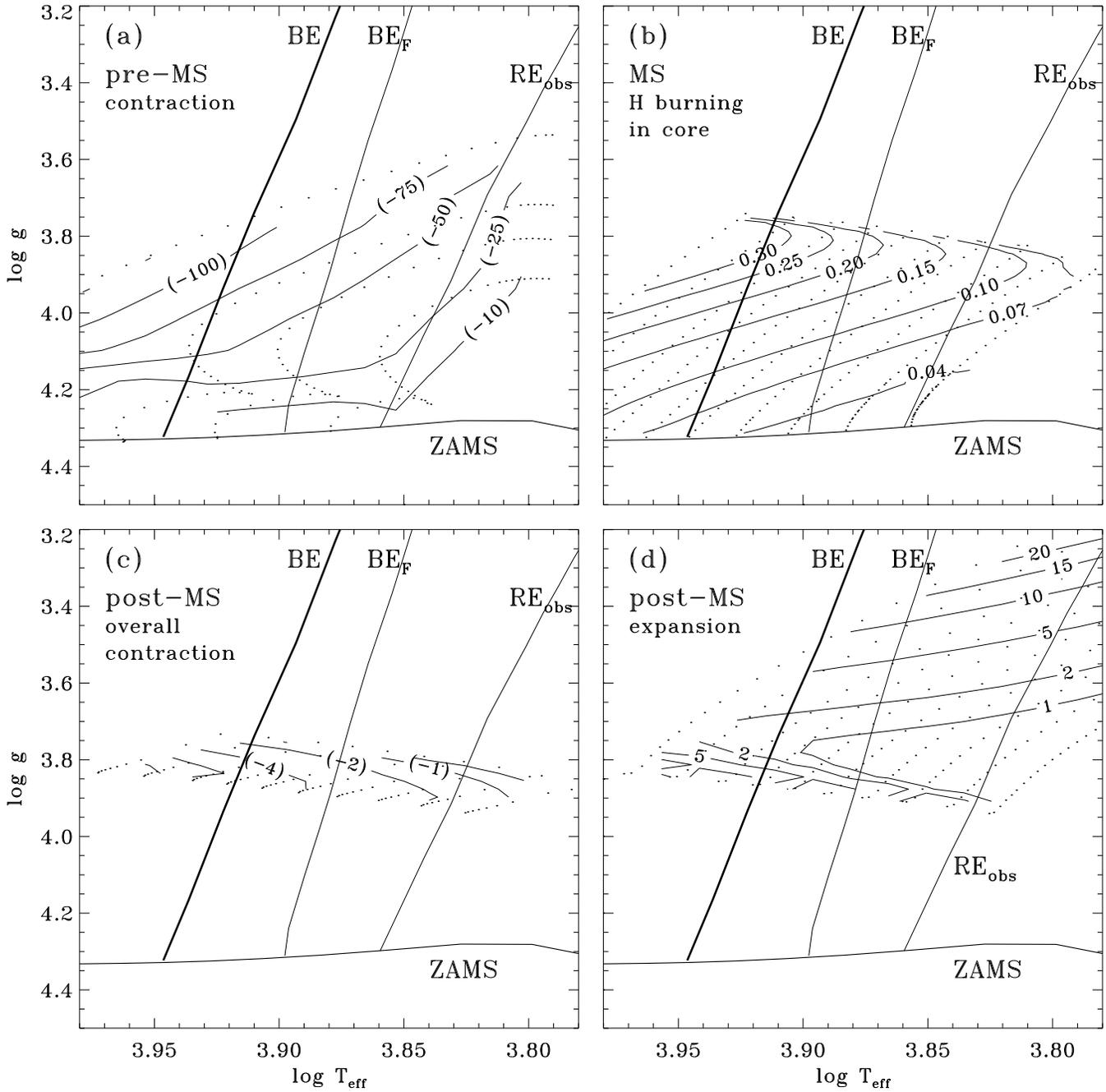}
\caption{Lines of constant period changes
in $\log g$ - $\log T_{\rm eff}$
diagram. The numbers give values of relative period changes,
${\rm (1/P_F) dP_F/dt}$, in units of $10^{-8}$ year$^{-1}$,
where ${\rm P_F}$ is
the period of the radial fundamental mode. The four panels correspond
to consecutive evolutionary stages:
(a) pre-MS contraction stage,
(b) MS stage of hydrogen burning in the stellar core,
(c) second overall contraction stage of effective hydrogen exhaustion
    in the core,
(d) post-MS expansion stage of the hydrogen burning shell source.
Points mark the position of the models in the corresponding
evolutionary
stages (compare with Fig. 1). The boundaries of the instability strip
are also shown with coding as in Fig. 1.}
\end{figure*}

The rates of evolutionary period changes were derived using computed
values
of the oscillation periods and the ages of the corresponding models.
The results are presented in Fig. 2 separately for
four consecutive evolutionary stages.

The periods of radial oscillations decrease during
the pre-MS and the second overall contraction stages,
and increase during the MS and post-MS expansion stages.
The rate of these changes is determined entirely by the rate of
evolution. The typical time spent by a star in the instability strip
is quite different for different evolutionary stages: depending on
stellar mass, this time is
about $0.001$--$0.01$ Gyr for the pre-MS stage,
$1$--$1.5$ Gyr for the MS stage,
$0.02$--$0.05$ Gyr for the second overall contraction stage
and $0.02$--$0.1$ Gyr for the post-MS expansion stage.
Therefore, values of period changes can differ one from another
by three orders of magnitude.
The fastest period changes, which can be easier detected,
occur during pre-MS evolution. Therefore, pre-MS variables seem to
be the best candidates to test the stellar evolution rate empirically.

\subsection{Effects of stellar rotation and convective overshooting}

Some indications in favor of significant overshooting
from convective cores of $\delta$ Scuti stars were
obtained from a comparison of evolutionary tracks with calibrated
photometric data (e. g. Napiwotzki et al. 1993).
An asteroseismological test, based on sensitivity of some nonradial
mode frequencies to the size of mixed stellar core, was proposed
by Dziembowski \& Pamyatnykh (1991).
Another important parameter which must influence stellar
evolution and pulsations, is rotation. Detailed studies of
non--evolutionary rotational effects
on the oscillation frequency spectrum are beyond the scope of
the present paper. Some estimates of the second order rotational
corrections to period ratios were obtained
by P\'erez Hern\'andez et al. (1994);
the complexity of the mode identification problem in the presence of
rotation is demonstrated by Pamyatnykh et al. (1997) in an attempt to
find a seismic solution to 13 observable frequencies of XX Pyxidis
(for a short discussion see Pikall et al. 1998). However, rotation
significantly modifies the structure and evolution of a star.
It may be desirable to compare the rate of period changes in such
models with that of standard non-rotating models.

To test the  effects of overshooting and rotation explicitly, we
computed two new families of evolutionary tracks of
1.5--2.5 $M_{\odot}$: models with overshooting from the
convective core and rotating models without overshooting.

The overshooting distance, $d_{\rm over}$, was chosen
to be $0.2{\cdot}H_{\rm p}$
where $H_{\rm p}$ is the local pressure scale height at the edge
of the convective core. A similar value of the overshooting parameter
was used by many authors (see, for example,
Schaller et al. 1992, Napiwotzki et al. 1993, Claret 1995).
The tracks with and without overshooting are shown in Fig. 3.

\begin{figure}
\epsfclipon
\epsfxsize=82mm
\epsffile{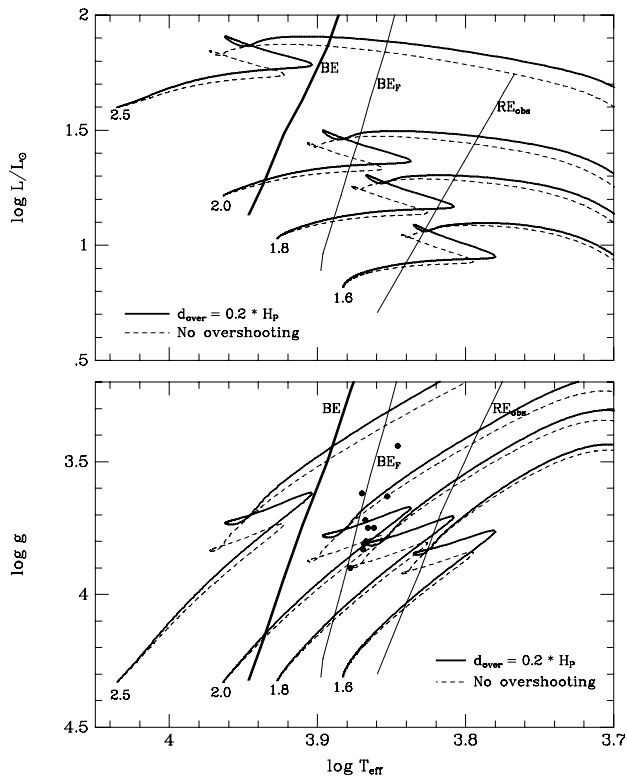}
\caption{Evolutionary tracks of 1.6, 1.8, 2.0 and 2.5 $M_{\odot}$
models with
(solid lines) and without convective overshooting (dashed lines).
The ZAMS models are practically identical in both cases because
they are chemically uniform.
The instability boundaries are the same as in Fig. 1.
In the lower panel, the location of 10 Pop.~I stars from Table~1 is
shown by filled circles. The values are based on the $uvby\beta$
data from the catalogue of Rodriguez et al. (1994) transformed to
$\log g$, $\log T_{\rm eff}$ using the
calibration of Moon \& Dworetsky (1985).
}
\end{figure}

The overshooting results in an extension of the MS--stage in the HR
Diagram due to an enlargement of the mixed core: more hydrogen fuel is
available for nuclear burning.
The displacement of the TAMS-points is about
$\Delta (\log T_{\rm eff}) = -0.02$,
$\Delta (\log g) = -0.1$ to $-0.15$\footnote
{
According to our computations with a given choice
of $d_{\rm over}$, overshooting influences the TAMS position 1.5--2
times smaller than according to Schaller et al. (1992), Napiwotzki
et al. (1993) or Claret (1995). The disagreement may be caused by
numerical effects as a test shows. To achieve a sufficient accuracy
in the oscillation computations we need to rely upon more detailed
stellar models than one uses usually in the evolutionary computations.
For example, our typical 2.0 $M_{\odot}$ MS model consists of
approximately 1300 layers, and there are 135 models between the ZAMS
and TAMS. If we increase both space and time step sizes by factor
of about 7--8, we can reproduce the corresponding evolutionary track of
Schaller et al. (1992) almost precisely.
}.
At a fixed effective temperature, a model with overshooting is slightly
more luminous than a standard model without overshooting.
Stellar lifetimes in the MS--stage are increased due to overshooting
by 12--14 percent.
At the TAMS, hydrogen is slightly more exhausted
in models with overshooting: the central hydrogen abundance is
about 4.0 \% (in mass) as compared with 4.5 \% in the case without
overshooting.

It is interesting to compare stellar lifetimes in different
evolutionary stages for the models with and without overshooting.
Let us consider segments of the evolutionary tracks of the
1.8 $M_{\odot}$ model
in the effective temperature range between the TAMS and the leftmost
point
of the second contraction stage. The star crosses that region three
times in its life on and beyond the MS. In the overshooting case
($\log T_{\rm eff}=3.8081-3.8671$, see Fig. 3)
these times are equal to 0.35, 0.034, and 0.019 Gyr for MS, second
contraction and post-MS expansion stages, correspondingly.
In the standard case without overshooting
($\log T_{\rm eff}=3.8260-3.8769$) these times are equal to 0.35,
0.038, and 0.056 Gyr. We see that in both
cases the second contraction times are similar and are one order
of magnitude shorter than in the final part of the MS--evolution.
On the other hand,
post-MS expansion in the overshooting case is a factor of 1.8
faster, and in the standard case a factor of 1.5 slower than the
corresponding second contraction. Such a significant difference in the
post-MS expansion
times between evolution with and without overshooting seems to be
important for statistical investigations of the $\delta$~Scuti stars.

For another family of models we assumed uniform (solid-body) stellar
rotation and conservation of global angular momentum during evolution
from the ZAMS. These assumptions were chosen due to their simplicity.
The same approach was used in our recent studies
of XX~Pyxidis (Handler et al. 1997a,b; Pamyatnykh et al. 1997)
The initial equatorial rotational velocity was assumed to be 150 km/s.
The tracks for rotating and non-rotating models are shown in Fig. 4.

\begin{figure}
\epsfclipon
\epsfxsize=82mm
\epsffile{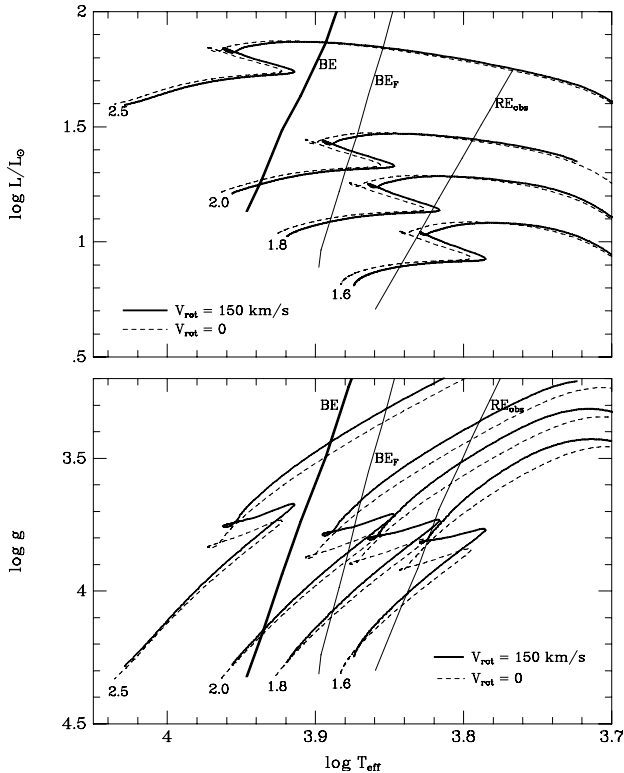}
\caption{Evolutionary tracks of uniformly rotating (solid lines) and 
non-rotating (dashed lines) stellar models 
with masses 1.6, 1.8, 2.0 and 2.5 $M_{\odot}$. (See text for details.)
The instability boundaries are the same as in Fig. 1.
}
\end{figure}

Rotation results in a shift of the tracks to smaller
$T_{\rm eff}$ and gravity. The displacement of the TAMS mimics
the overshooting effect, with
$\Delta (\log T_{\rm eff})$, $\Delta (\log g)$ smaller
by a factor 1.5 in comparison with those for the overshooting case
(compare Figs. 3 and 4).
The tracks of rotating models lie slightly above the tracks
of non-rotating models in $\log g - \log T_{\rm eff}$ diagram,
and below the corresponding tracks in
$\log L - \log T_{\rm eff}$ diagram which can be easily
explained by decreasing effective gravity due to centrifugal force.
The MS lifetime for rotating models is only 0.5--1.0 percent
longer than that for non-rotating models.
With our assumption on the global angular momentum conservation
the equatorial rotational velocity is decreasing during the
MS--evolution from 150 km/s at the ZAMS to about 120 km/s at the TAMS.

On the MS, the effects of the overshooting and
rotation on period changes are determined
practically entirely by mentioned changes in positions of the tracks.
In a plot similar to Fig. 2b, lines of constant period changes for
the overshooting case will be prolongated to smaller gravities due to
the TAMS displacement. For the rotating models we will obtain
a small shift of the isolines to lower gravities according
to both the ZAMS and the TAMS displacements.
We disregard the rotation effects on evolutionary period changes
in the rest of this paper because they mimic the overshooting
effects but are less pronounced than these.

\subsection{Period changes during fast phases of post-MS evolution}

We will show in this subsection that fast period changes,
more pronounced for the overshooting case, can
occur during very short phases of the post-MS evolution.
As it can be seen from Fig. 3,
there is a noticeable difference in the morphology of the tracks
with and without overshooting at the end the second overall
contraction stage. Here a negative and
a positive peak in the period changes are produced. Physically, at this
stage a significant reorganization of stellar interior structure occurs:
towards the end of the second overall contraction stage
the convective core begins
to decrease very quickly due to progressive inefficiency of the
nuclear-energy generation in the center, which, in turn, is caused by
hydrogen exhaustion. This process occurs more suddenly in the
overshooting case: for example, in the 1.8 $M_{\odot}$ model the
decrease of the convective
core from the value of 0.005 (mass fraction) to 0.0 lasts
for about 0.9 million years
which is three times faster than in the standard case.
At the end of the overall contraction stage
the convective core disappears\footnote
{
Of course, convective overshooting also
disappears, but we keep the term `model with overshooting' or
`the overshooting case' for later evolutionary stages to distinguish
this case from the standard one.
}.
As hydrogen is effectively exhausted in the center, this region
rapidly cools and contracts. A significant part of the luminosity is
provided with release of thermal and gravitational energy.
Simultaneously
with the cooling at the center (in the inner 6--7 \% of stellar mass)
the heating outside this region occurs where hydrogen is more abundant.
As a consequence, a nuclear hydrogen-burning shell is established
outside the hydrogen-exchausted core. The stellar luminosity adjusts
quickly to the energy production by the nuclear shell-burning source.
In the HR diagram, such `non-explosive hydrogen ignition' occurs on the
ascending part of the evolutionary track soon beyond a small loop.

Details of post-MS stages of standard stellar evolution were described
more than thirty years ago by Iben (1967 and references therein).
When discussing a very fast stage of the convective core disappearance
and of the hydrogen-burning shell development,
the overshooting {\it versus} non-overshooting analysis seems to be
similar to the analysis of the standard evolution of models of different
mass: for a higher mass we proceed to construct
the models with more extended convective cores.
Therefore, as was noted by Iben (1966) in his discussion
of the $5 M_{\odot}$ {\it versus} $3 M_{\odot}$ evolution,
the nuclear fuel disappears suddenly over a somewhat larger
fraction of the interior. As a consequence, `semidynamic' effects
of the stellar structure reorganization are manifestated more
clearly (see small loops in our tracks).

The time spent by a 1.8 $M_{\odot}$ star in the small loop near
the end of
overall contraction stage is equal to 0.0015 and 0.0014 Gyr for
the overshooting and the standard case, correspondingly. It is
25 times shorter than the total time of overall contraction.

\begin{figure}
\epsfclipon
\epsfxsize=82mm
\epsffile{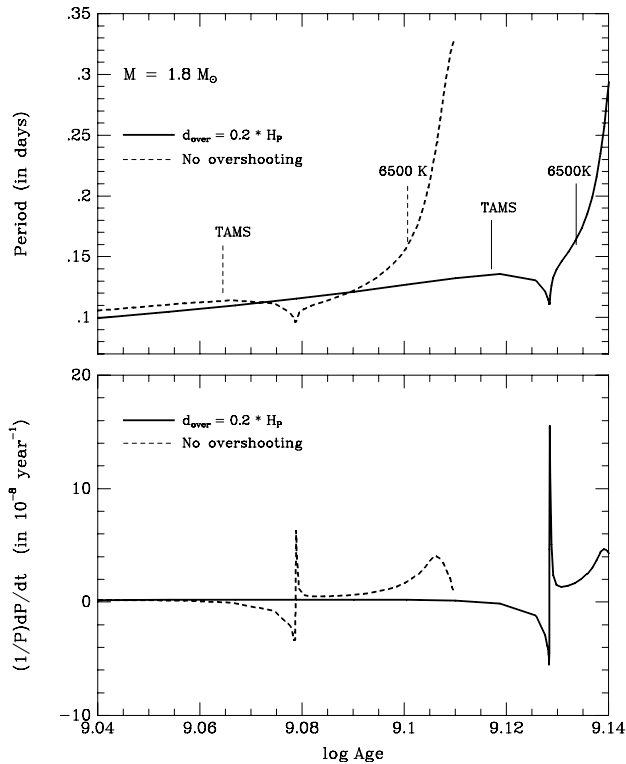}
\caption
{Period of radial fundamental mode and relative period change for
late MS and post-MS (right of the TAMS) evolution of 1.8 $M_{\odot}$
model with and without overshooting. Short vertical lines in the upper
panel mark ages of the TAMS models and
those of the models with $T_{\rm eff}$ = 6500~K which is close to the
Red Edge of the instability strip (see Fig. 3).
}
\end{figure}

The periods of the radial fundamental mode and their changes during
late MS and post-MS evolution of a 1.8 $M_{\odot}$ star are shown
in Fig. 5.
Note that the MS--evolution with overshooting lasts for about 13~\%
longer than in the standard case (1.31 and 1.16 Gyr, correspondingly).
At the same time, post-MS evolution occurs more rapidly
in the overshooting case. The dips in the period behaviour and
corresponding peaks in the rate of the
period change are seen clearly\footnote
{
The lower panel of Fig. 5 is similar to plots
presented by Heller \& Kawaler (1988) who studied
evolutionary period changes in rapidly oscillating Ap stars.
We compared our standard results (without overshooting) for
1.6, 2.0 and 2.5 $M_{\odot}$
with those of Heller \& Kawaler and found agreement in the MS stage.
However, our peak values of period changes are three times larger
for negative peaks and six times larger for positive peaks.
Also, the positions of our peaks are shifted to higher ages.
These differences can be explained by differences in the evolutionary
tracks. Somewhat higher new OPAL opacities in the interior
result in an enlargement of the convective core and in a corresponding
extension of MS evolutionary stage. The situation is similar to our
comparison of the overshooting case with the standard case
(see Fig. 5).}

These features are produced during a
very short phase around the end of overall contraction (small loop
in the evolutionary track, see Fig. 3).
In the overshooting case, the maximum positive period change is about
$16\cdot10^{-8}$ yr$^{-1}$ which is 2.5 times larger than that in the
standard case. We stress that due to the short duration of this stage
the probability to observe these large period changes is small.

\section{Observed period changes}

We have collected, examined, and in some cases recalculated the
observed period changes for $\delta$ Scuti stars. Most available
data refer to stars with radial pulsation and large amplitudes.
The period changes are usually derived from a collection of observed
times of maximum light. Regrettably, a potentially confusing variety
of different units can be found in the literature. In a few cases,
these have lead to errors. Because of these problems
it appears prudent to list the equations and units used in this paper:

Let (O-C) refer to the difference in the observed and computed times of
maxima measured in days. If the adopted period, P, is correct and
refers to the beginning of a long series of observations covering a
time t (also in days), then
\[(O-C) = 0.5 ~\frac{1}{P} ~\frac{dP}{dt}~t^2.\]
This equation is similar to an expression derived by
Percy et al. (1980), where phase units are used. It is customary
to express (1/P)dP/dt in units of year$^{-1}$, so that the value of
(1/P)dP/dt found above has to be multiplied by a factor of 365.25.

Ambiguities should not be present if the observations are expressed
in terms of the times of maximum light (epoch):
\[T_{\rm max} = T_o + P_o~N + 0.5~\beta~N^2,\]
where $\beta$ = P dP/dt, T${_o}$ refers to the epoch and P${_o}$ to
the period at the time of the listed epoch.


\begin{table*}
\caption[]{Observed period changes of Delta Scuti stars}
\begin{tabular}{llcll}
\hline
\noalign{\smallskip}
Star & Period & (1/P)dP/dt & References & Comments \\
 & (days) & (year$^{-1}$ x 10$^{-8}$) \\
\noalign{\smallskip}
\hline
\\
\multicolumn{5}{l}{\hspace{-7pt}Population I stars, radial pulsation}\\
\\
IP Vir & 0.067 & -0.5 & Hintz (priv. comm.) & no jump, Note\\
GP And & 0.079 & 13 & Rodriguez et al. 1993a\\
AE UMa & 0.086 & -48 & Hintz, Hintz \& Joner (1997)\\
EH Lib & 0.088 & small & Mahdy \& Szeidl (1980)\\
 & & & Yang et al. 1992\\
BE Lyn = HD 79889 & 0.096 & ... & Kiss \& Szatmary (1995), & binary
light-time effects?\\
& & & Rodriguez et al. (1995)& Note\\
YZ Boo & 0.104 & 3 & Hamdy et al. (1986)\\
 & & & Jiang (1985)\\
AI Vel & 0.112 & 0 & Walraven et al. (1992)\\
 & 0.086 & 15 & & Note\\
SZ Lyn & 0.121 & 7 & Moffett et al. (1988), Papar\`{o} et al. (1988)&
Note \\
AD CMi & 0.123 & 4 & Fu \& Jiang (1996), Rodriguez et al. (1995) &
Note\\
RS Gru & 0.147 & -11 & Rodriguez et al. (1995)\\
DY Her & 0.149 & -4  & Szeidl \& Mahdy (1981), Yang et al. (1993)\\
VZ Cnc & 0.178 & ... & Arellano Ferro, Nu\~{n}ez \& Avila (1994)&
Note\\
BS Aqr & 0.198 & -5 & Yang et al. (1993)\\
\\
\multicolumn{5}{l}{\hspace{-7pt}Population I stars, nonradial
pulsation}\\
\\
XX Pyx & 0.026 & 340 & Handler et al. (1997a) & jump possible\\
 & 0.028 & - 910 & & jump possible \\
 & 0.030 & small \\
$\tau$ Peg & 0.054 & variable & Breger (1991)& Note\\
4 CVn & 0.116 & -110 & Breger (1990b) & jump possible\\
 & 0.171 & -300 & & jump possible\\
\\
\multicolumn{5}{l}{\hspace{-7pt}Population II stars = SX Phe stars,
radial pulsation}\\
\\
BL Cam & 0.039 & 29 & Hintz et al. (1997)& Note, no jump\\
SX Phe & 0.055 & -2 & Thompson \& Coates (1991), Coates et al. (1982)&
jumps likely\\
 & 0.043 & -16 \\
CY Aqr & 0.061 & ... & Powell et al. (1995) & 4 jumps\\
show diagram\\
DY Peg & 0.073 & - 3 & Mahdy (1987), Pe\~{n}a et al. (1987)\\
XX Cyg & 0.135 & positive & Szeidl \& Mahdy (1981) & jump in 1942 \\
\\
\hline
\\
\end{tabular}
IP Vir: Pop. I nature suspected (Landolt 1990).

BE Lyn: A parabolic fit would lead to
(1/P)dP/dt = 40 x 10$^{-8}$ year$^{-1}$ (Liu \& Jiang 1994,
Rodriguez et al. 1995), but the scatter is large. The
binary hypothesis by Kiss \& Szatmary (1995) with an orbital
period $\sim$ 2350 d looks promising, but needs to be confirmed.

AI Vel: The value of dP$_1$/dt given by Walraven et al. (1992) has been
multiplied
by P$_1^2$ to correct for an error. The new value can be confirmed by
inspection of Fig. 1 of that paper.

SZ Lyn: Light-time corrections for the binary orbit have been applied.

AD CMi: Fu \& Jiang (1996) list an additional explanation of the stage
shifts in terms of orbital light-time effects.

VZ Cnc: Data cannot reliably distinguish between constant period,
abrupt period changes or orbital light-time effects. Some additional
data by Arellano Ferro, Avila \& Gonzalez (1994) are available.

$\tau$ Peg: Period increases and decreases at different times,
jumps possible.

BL Cam: Fig. 4 and equation 3 of Hintz et al. (1997) lead to a period
change of 0.02 s in the last 20 years, not the value of 0.009 s listed
in the paper.

\end{table*}

Table 1 lists a summary with some of the recent references and can be
regarded as an update of earlier collections given in Breger (1990a)
and Rodriguez et al. (1995). We have deliberately not listed the
statistical uncertainties of the dP/dt values, because in almost
all cases the published or calculated uncertainties are
unrealistically small. This can be demonstrated for a number of stars
for which new observations have become available and new values of the
standard deviation were computed. However, we have omitted
the few stars for which we regard the period changes as very uncertain.
Although
not included in this paper, the calculation of reliable standard
deviations as well as upper and lower limits for those stars with
small period changes are important.

The star 28 And was not included in Table 1, because
Rodriguez et al. (1993b) have
found that the times of maxima can be fit by a constant period,
i.e. the star's period changes are smaller than the uncertainties
of determination. Furthermore,
the period changes of KZ Hya = HD 94033 (Hobart et al. 1985) are
regarded as too uncertain to be included here.

\section{Period changes among Pop. I $\delta$ Scuti stars}

\begin{figure}
\epsfclipon
\epsfxsize=85mm
\epsffile{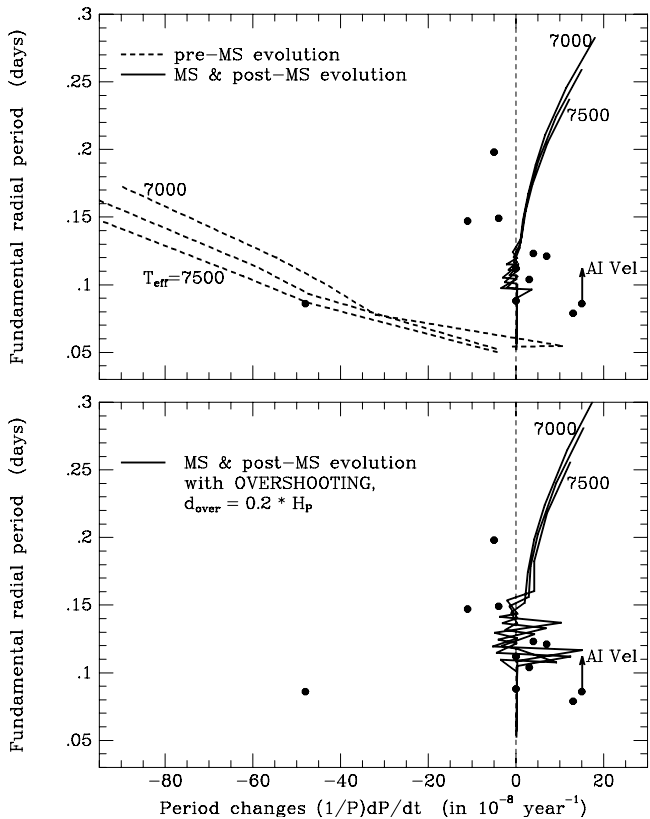}
\caption
{Comparison between the observed (points) and
computed (curves) period changes of Pop. I
radial pulsators. The arrow corresponds to the observed period change
of the radial first overtone
mode of AI Vel corrected to the value of the fundamental mode.
The theoretical data for models
with (lower panel) and without (upper panel)
core overshooting are given for three effective temperatures.
The zigzag behaviour of the curves at periods of about 0.1--0.15 days
corresponds to the very fast evolution of models of different masses
at the end of the second overall contraction stage.
During this phase sudden period changes occur (see Fig. 5).
The inclusion of the overshooting extends the
domain of zigzag behavior.}
\end{figure}

In Fig. 6 we compare the observed period changes of the Pop. I radial
pulsators with the values expected from stellar evolution calculations
described in a previous section of this paper.
The stars with known period changes are generally those
with high amplitudes (HADS), which are restricted to the
central part of the instability strip. Consequently, the comparison
occurs in a limited temperature range of about 7000 to 7500 K. For the
figure we have chosen the size of the radial fundamental period as the
indicator of the evolutionary state. For these radial pulsators, this
is a better indicator than the position in the HR Diagram,
because of the uncertainties of about $\pm$ 0.3 mag in the photometric
absolute magnitude calibrations. The uncertainties in the values of the
measured periods are negligible.

Fig. 6 also shows the computed period changes for three temperatures of
7000, 7250 and 7500 K. Models with and without
overshooting were considered. Not surprisingly, the rate of
evolution is not a strong
function of temperature. Furthermore, in the case of standard
evolution without overshooting, only for the pre-MS and most evolved
stars does the predicted rate of period change
exceed (1/P)dP/dt = $\pm$ 10$^{-7}$ year$^{-1}$.
Taking overshooting into account results in large period changes also
at the end of the second overall contraction stage.

The expectation of generally positive period changes increasing in size
with increasing evolutionary status is not borne out by the
observations. In fact, the stars
divide equally into groups with increasing and decreasing periods.
Furthermore, the observed period changes are generally one order of
magnitude
larger than expected. The disagreement should not be caused
by incorrectly computed rates of stellar evolution, since
the situation will
not be changed significantly  if we take into account some additional
effects such as rotation or convective overshooting.

An exception may only occur during the very fast stages of post-MS
evolution as has been discussed in subsection 2.4.
It can be seen in Fig. 6 that overshooting results in a
significant enlargement of the zigzag domain and a shift
to somewhat longer periods. In principle, up to half of observed
period changes could now be accounted for.
Let us examine a possible contender for the zigzag domain in more
detail: the star AI Vel. The values of both the observed radial
fundamental and first overtone modes place the star inside or near
the zigzag domain.
The value of the period ratio is accurately known and can be fitted
to this fast evolutionary stage without any problems.
This is illustrated in Fig. 7, which is similar to Fig. 11 in the
paper by Petersen \& Christensen-Dalsgaard (1996). It can be seen
that a model of 2.0~$M_{\odot}$ with the overshooting parameter
$d_{\rm over}$ between 0.0 and $0.2{\cdot}H_{\rm p}$ can fit the
observational periods of AI~Vel very precisely.

\begin{figure}
\epsfclipon
\epsfxsize=82mm
\epsffile{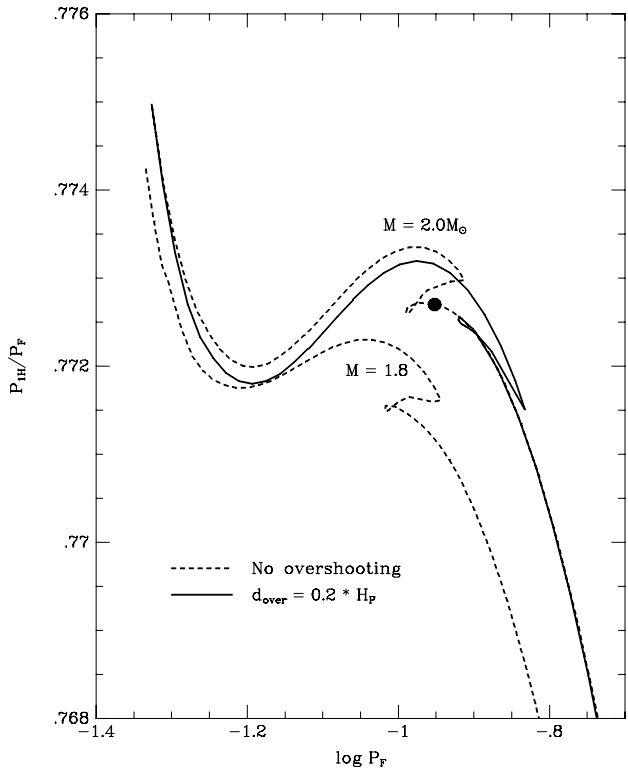}
\caption
{Period ratio diagram for 2.0~$M_{\odot}$ evolutionary models with
and without overshooting. The leftmost points correspond to the ZAMS
models. For comparison, the standard 1.8~$M_{\odot}$ track
is shown. The filled circle marks the position of AI~Vel with data
taken from Petersen \& Christensen--Dalsgaard (1996).}
\end{figure}

Consequently, AI Vel may be an example of a star in the zigzag domain
of Fig. 6 so that
the observed period change of the first overtone could be explained.
However, the radial
fundamental mode of AI Vel is also observed and shows no period
changes. This is not in agreement with the computations\footnote
{
There exists a very simple test to check whether observed period
changes can be explained by the evolutionary effects. If a star
shows multiperiodic behaviour and if the extracted frequencies
can be interpreted as pure acoustic modes (for example, the radial
fundamental mode and the first overtone in most of double-mode
variables), we expect the logarithmic rate of period changes to be
the same for all modes because their periods are determined almost
entirely by the mean stellar density.}.
Furthermore, the agreement between positions of many stars in Table 1
with the zigzag domain may be accidental. Stellar evolution in this
region around the end of the second overall contraction stage is very
fast, as was already
demonstrated in Fig. 5: for a 1.8 $M_{\odot}$ star, the stage of the
negative and positive
peaks in the period change takes not longer than 1.5 million years.
The probability of
finding such a high fraction of the observed stars in this short
evolutionary stage is essentially zero. Consequently, the overall
disagreement between the observed and expected period changes is
not removed.

Among the Pop. I radially pulsating $\delta$ Scuti stars,
AE UMa has the fastest measured period
change of --5 x 10$^{-7}$ year$^{-1}$. The data look reliable.
The rate of the period decrease is similar to that expected for
pre-MS variables (see Fig. 6).
However, there exists no evidence that AE UMa is a pre-MS star.

In an earlier discussion of three giants of similar luminosity and
temperature, Breger (1990a) noted that, at least in principle, the
theoretical evolutionary tracks could be adjusted to lead to
decreasing radii in this part of the HR Diagram. However, the present
analysis also shows disagreements for the
less evolved stars. This makes an explanation in terms of completely
erroneous evolutionary tracks improbable. An additional, but possibly
unneeded argument concerns
the observed period--luminosity relation shown by the hundreds of
$\delta$~Scuti detected so far,
indicating a radius increase for the more evolved stars with longer
periods (see Introduction).

The disagreement between observed and computed evolutionary period
changes is even worse for the stars showing nonradial pulsation.
However, the calculation of the nonradial pulsation
frequencies and their evolutionary changes is more complicated than
for radial pulsation. The reason is that the rotational splitting may
change with evolution so that modes with different
values of the azimuthal order, $m$, may show different period changes,
even in the same star (as observed).

We conclude that the period changes observed
in \mbox{Pop. I} $\delta$ Scuti stars are not caused by stellar
evolutionary changes. The period changes cannot at this
stage be used to check stellar evolution calculations. Of course,
the evolutionary changes should still be present in these stars,
but are masked by other effects causing period changes. A much
larger sample of stars is needed to average out these other effects.

\subsection{On non--evolutionary period changes}

The physical origin of the observed nonevolutionary period changes is
not known. There are several suggestions in the literature:

Stellar companions can produce (O-C) shifts due to the light-time
effects caused by the orbital motion.
With limited amounts of data the observed shifts can often be matched
equally well by both parabolic and trigonometric functions, so that
a binary origin of the time shifts cannot be ruled out. The ambiguity
has been demonstrated by Fu \& Jiang (1996) for the star AD CMi.
The situation is, however, not hopeless: the orbital changes reverse
themselves over a longer time scale.

It is possible that most of period changes which are observed
in $\delta$ Scuti stars are caused by nonlinear mode interactions,
not by some specific processes in the interiors like considered
by Sweigart \& Renzini (1979) for RR Lyrae models almost twenty years
ago. (The authors proposed random mixing events in a semiconvective
zone to be the origin of period changes of both signs, especially
sudden period jumps. It was noted by Rodriguez et al. (1995) and
others that some of the changes in $\delta$ Scuti stars resemble
those also seen in RR Lyrae stars. However, there is no semiconvective
zone in the less evolved $\delta$ Scuti stars.)

The effect of period changes due to nonlinear mode interactions was
demonstrated
in a completely different context by Buchler et al. (1995).
The authors studied nonlinear behaviour of rotationally split
dipole ($\ell=1$) modes in pulsating stars.
In particular, when all three components of the triplet are excited
and the amplitudes are constant, the direct nonlinear interactions
between components result in symmetric frequency spacing.
This is contrary to the results from linear theory, in which the
second order effects of rotation cause the rotational mode
splitting to be highly nonequidistant (see Dziembowski \& Goode, 1992).

Another specific case of the nonlinear mode interactions was studied
by Moskalik (1985). He considered the problem of
resonant coupling of an unstable mode to two lower
frequency stable modes. For $\delta$ Scuti stars
such an approach means that a low--order radial or nonradial unstable
mode interacts with two stable gravity modes.
The coupling can result in a periodic amplitude modulation.
The period of the modulation depends mainly on the
excitation rate of the unstable mode and is supposed to be of the order
of years in $\delta$ Scuti stars. During most of the modulation cycle
the amplitude of the unstable mode grows nearly exponentially but
finally decreases rapidly due to nonlinear interactions with the two
stable modes. The large jump of pulsation phase occurs
simultaneously -- positive or negative
depending on the sign of the frequency mismatch in the resonance
condition. Before and after this jump one can observe very fast changes
of the pulsation period.
For $\delta$ Scuti models in a wide range of oscillation
parameters the rate of these fast period changes (1/P) dP/dt
was found to lie, approximately, between
$7.0\cdot10^{-4}$ and $1.0\cdot10^{-3}$ year$^{-1}$. These large
values can be appropriate to interpret some changes like observed
in XX~Pyxidis (see Handler et al. 1997a).
As mentioned by Moskalik, a very serious limitation of his
study is connected with the assumption that only three modes take part
in the interaction.

If the radially pulsating Pop. I stars with observed period changes
are plotted on the $\log g$ -- $\log T_{\rm eff}$ diagram (Fig. 3),
it can be seen that all stars are located
close to the theoretical Blue Edge of the fundamental radial mode. This
conformity may be accidental because the fundamental Blue Edge lies
approximately along the most populated central part of the instability
strip. Moreover, the calibration errors may be significant too. But if
this agreement is real, then, potentially, it can take a part in the
future period change interpretations. At this line the fundamental
radial mode instability appears or disappears depending on the
direction of evolution. That means that quite significant changes
in the excitation or damping rate of different modes may occur.
They must influence the efficiency of the nonlinear interactions
between different modes which, in turn, may influence the periods
more significantly than during stages of more steady oscillations.

The situation for comparing observed period changes with stellar
evolution calculations may be brighter for pre-MS stars. The
calculations shown in Fig. 6 indicate expected period changes a
factor of 10 to 100 times larger than for MS stars.
In fact, these expected rates of period change are even larger than
the nonevolutionary period changes of MS stars.
Consequently, the evolutionary changes might not be hidden among
other effects. A few $\delta$ Scuti stars with probable pre-MS status
are known, e. g. the stars W2 and W20 in NGC 2264. So far, the
variability of these stars has not been studied since their
discovery as $\delta$ Scuti stars (Breger 1972).
The study of period changes among the pre-MS stars appears to be very
promising, in contrast to MS and post-MS stars.

\section{Observed period changes in SX Phe stars}

The evidence is accumulating that most of the observed period changes
in the Pop. II $\delta$ Scuti stars occur in sudden jumps, followed
by constant or nearly-constant periods. This
is clearly shown in the star CY Aqr (Powell et al. 1995). Furthermore,
the abrupt period change of 1989 reversed the decreases in period from
the previous two period changes. This convincingly argues against the
interpretation of the period changes
in terms of long-term stellar evolution (at least in this star).

Other stars tend to show a similar behavior: The star XX Cyg had a
sudden period change in 1942 (Szeidl \& Mahdy 1991), while the
(O-C) variations of SX Phe are best interpreted as two sudden jumps
in 1969 and 1975 (Thompson \& Coates 1991). In all three stars, the
jumps were of the order of $\Delta P/P \sim$ 10$^{-6}$.

The data for the star DY Peg can be interpreted equally well as either
a sudden jump in 1961
or a continuous change (see Mahdy \& Szeidl 1980). On the other hand,
BL Cam shows a continuously changing period.

The character of the observed period changes in $\delta$ Sct stars
suggest that they are caused by nonlinear
effects in pulsation and not by stellar evolution, at least in
most cases. Therefore the main application of the data we collected
here could be used in the future to test the theory of
nonlinear multimode stellar oscillation.

\acknowledgements
Part of the investigation has been supported by the
Austrian Fonds zur F\"{o}rderung der wissenschaftlichen Forschung,
project number S7304. AAP acknowledges also partial financial
support of the computations from
the Polish Committee for Scientific Research (grant 2-P304-013-07)
and from the Russian Fund for Basic Research (grant 95-02-06359).
The authors are also indebted to Wojciech~Dziembowski
and Pawe{\l}~Moskalik for enlightening discussions.

\end{document}